\documentclass[preprint,12pt]{elsarticle}

\usepackage{graphicx}
\usepackage{amssymb}

\journal{Nuclear Inst. and Methods, A}

\begin{document}

\begin{frontmatter}

\title{Proposal for measuring the quantum states of neutrons in the gravitational field with a CCD-based pixel sensor}

\author[tohoku]{T. Sanuki\corref{cor1}}
\ead{sanuki@awa.tohoku.ac.jp}
\author[tokyo,icepp]{S. Komamiya}
\author[tokyo]{S. Kawasaki}
\author[tokyo]{S. Sonoda}

\cortext[cor1]{Corresponding author. Tel.:+81-22-795-6727; fax:+81-22-795-6728}

\address[tohoku]{Graduate School of Science, Tohoku University, 
Sendai-shi, Miyagi 980-8578, Japan}
\address[tokyo]{Graduate School of Science, The University of Tokyo, 
Bunkyo-ku, Tokyo 113-0033, Japan}
\address[icepp]{International Center for Elementary Particle Physics, The University of Tokyo, 
Bunkyo-ku, Tokyo 113-0033, Japan}

\begin{abstract}
An experimental setup is proposed
 for the precise measurement of the quantum states of ultracold neutrons
 bound in the earth's gravitational field.
The experiment utilizes a CCD-based pixel sensor
 and magnification system to observe the fine structure
 of the neutron distribution.
In this work,
 we analyzed the sensor's deposited energy measurement capability and found that its
 spatial resolution was $5.3\,\mu\mathrm{m}$.
A magnifying power of
 two orders of magnitude
 was realized by using a cylindrical rod as a convex mirror.
\end{abstract}


\end{frontmatter}

\section{Introduction} 
\label{sec:intro}
The energy of a neutron is quantized
 when it is bound in the earth's gravitational field.
As a consequence, the vertical distribution will not be smooth;
 rather, the distribution will exhibit modulation.
The scale of this density modulation is calculated to be 
 $(\hbar^2/2m_n^2g)^{1/3} \sim$ $6\,\mu$m,
 where $m_n$ is the neutron mass.
Such quantum states have recently been observed in 
 \cite{Nesvizhevsky:2002ef,Nesvizhevsky:2003ww,Nesvizhevsky:2005ss}.
A more precise measurement of these quantum states can be used
 to verify the validity of the gravitational equivalence principle and
 search for new short-range forces.

In the previous experiment,
 the vertical spatial density of neutrons was observed
 by using a uranium-coated plastic nuclear track detector (CR39).
The detector's spatial resolution was estimated to be as high as $\sim 2\,\mu$m
 by observing the distribution of neutrons \cite{Nesvizhevsky:2005ss}.
Since temporal information is not recorded in such plastic nuclear track detectors,
 these detectors are intolerable to accidental events and background signals.
We believe
 that a pixel detector with a real-time readout is an ideal device
 for precisely measuring positions in applications with low event rates. 

Silicon pixel sensors, such as CCDs, can measure
 both the spatial and temporal position of charged particles
 as the particles pass through the sensor. 
With the addition of a coating
 that converts neutron events into charged particle events
 that may be observed with the CCD, 
 the position of the original neutron event
 can be determined
 very precisely and in almost real time. 
We have developed such a ``fine-pixel neutron detector,''
 which is based on commercially available CCD technology.

In order to observe the structure of the neutron density modulation
 at scales much smaller than the size of a CCD pixel,
 we have designed and demonstrated a simple magnification system.
Our system uses a cylindrical rod as a convex mirror
 that can stretch the small modulation pattern of the neutron distribution
 by a factor of 40 or more.

In the following we propose an experimental setup that may be used to precisely measure
 the quantum states of ultracold neutrons
 bound in the earth's gravitational field.

\section{CCD-based neutron detector}
\label{sec:ccd}

\subsection{CCD sensor}
\label{subsec:ccd}
In order to detect neutrons with a CCD sensor,
 the sensor needs to be coated with a neutron converter, like $\mathrm{^6Li}$,
 which produces charged particles via the following process:
 \begin{equation}
 \mathrm{^6Li + n \rightarrow \alpha(2.05MeV) + T(2.73MeV)}.
 \label{eq:6Li}
 \end{equation}
The resulting alpha particle and triton can penetrate through silicon
 to characteristic depths of $8\,\mu\mathrm{m}$ and $40\,\mu\mathrm{m}$, 
 respectively.
If there is thick, insensitive material in front of the CCD sensitive volume
 the charged particles that are produced will not be able to enter the sensitive volume.
Therefore, back-thinned CCD is ideal for neutron detection.
In our study a 
 S7170-0909 (Hamamatsu Photonics K.K.) was selected
 for its wide active area and pixel size.
Other specifications of this CCD are listed in Table\,\ref{tbl:spec}.
\begin{table}[htdp]
\caption{Specifications of the CCD S7170-0909 ($\mathrm{T_a=25^\circ C}$).}
\begin{center}
\begin{tabular}{ll}
\hline
Parameter & \\
\hline
Active Area	&
12.288 $\times$ 12.288 mm
\\
Number of Pixels	&
512 $\times$ 512
\\
Pixel Size	&
24 $\times$ 24 $\mu$m
\\
Frame Rate	&
0.9 frames/s
\\
Spectral Response Range		&
200 to 1100 nm
\\
Full Well Capacity (Vertical)	&
300 k$e^-$
\\
Dark Current Max. $\mathrm{0^\circ C}$	&
600 $e^-$/pixel/s
\\
Readout Noise	&
8 $e^-$rms
\\
\hline
\end{tabular}
\end{center}
\label{tbl:spec}
\end{table}%

\subsection{Neutron converter}
\label{subsec:conv}
Since the secondary charged particles are emitted isotropically
 from the neutron conversion point,
 these particles can be produced
 at a large incident angle with respect to the CCD surface.
For events of this kind,
 the position detected is shifted from the position where the particle is emitted,
 reducing the spatial resolution of the detector
 if there is a gap between the CCD and the neutron conversion point.
In order to determine a neutron's incident position precisely,
 the neutron-converter layer should be very thin 
 and should be attached directly to the CCD
 so that the secondary charged particles enter the CCD
 immediately after the nuclear reaction.

In our design, 
 the neutron converter is formed directly on the CCD
 and the thickness is much less than the size of the CCD pixels $24\,\mu\mathrm{m}$.
The converter was processed
 using the vacuum evaporation facility
 at the Kyoto University Research Reactor Institute (KURRI)
 in Japan \cite{Ebisawa:1994ru,KUR}.
A $0.09\,\mu\mathrm{g/cm}^2$ ($20\,\mathrm{nm}$) Ti layer was formed on the CCD,
 providing an adhesive surface.
A $\mathrm{^6Li}$ layer of $0.11\,\mu\mathrm{g/cm}^2$ ($230\,\mathrm{nm}$) was then formed
 on the Ti layer
 to function as a neutron converter.
The $\mathrm{^6Li}$ converter was covered
 with a $0.09\,\mu\mathrm{g/cm}^2$ ($20\,\mathrm{nm}$) Ti layer
 to prevent the $\mathrm{^6Li}$ layer
 from reacting with moisture in the atmosphere.

\subsection{Energy measurement}
\label{subsec:energy}
In order to investigate the feasibility of using a CCD for neutron detection,
 the CCD coated with a neutron converter was irradiated with cold neutrons
 at the MINE2 beam line of the research reactor JRR-3M 
 of the Japan Atomic Energy Agency (JAEA); the neutron wavelength was $0.88\,\mathrm{nm}$.
The detection efficiency for cold neutrons was measured to be 0.3~\%,
 consistent with the efficiency estimated
 from a cross-section of the nuclear reaction (\ref{eq:6Li})
 and the thickness of the $\mathrm{^6Li}$ converter.
The efficiency corresponds to 13~\% for ultracold neutrons,
 where the cross-section is inversely proportional to velocity.
To detect ultracold neutrons more efficiently,
 a thicker $\mathrm{^6Li}$ layer is required in the neutron converter.
 
\begin{figure}[htbp]
\begin{center}
\includegraphics[width=0.9\columnwidth]{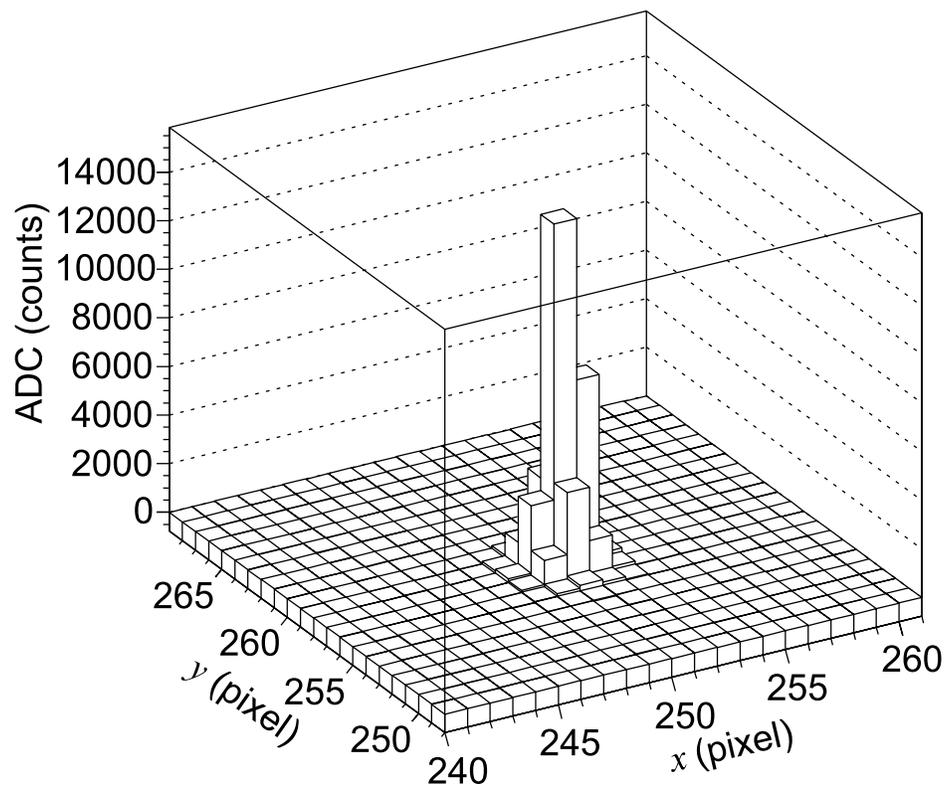}
\caption{Typical observed event.}
\label{fig:cluster}
\end{center}
\end{figure}

Fig.\,\ref{fig:cluster} shows an example of charge distribution in the CCD pixels for one event. 
A secondary charged particle loses its energy inside the CCD
 and creates as many as $10^6$ electron-hole pairs.
Due to diffusion inside the silicon,
 the electrons created during ionization are distributed in neighboring pixels to form a cluster.

The total charge inside the cluster is proportional to the energy deposited by the incoming particle. 
A correlation between the total charge and the total energy deposited was calibrated
 using an alpha source $\mathrm{^{241}Am}$.
The alpha source and the CCD were first placed inside a vacuum chamber;
the distance between them was $122.6\,\mathrm{mm}$.
Then the chamber was filled with dry $\mathrm{N_2}$ gas
 and the pressure varied between
 $3.0\times10^4\,\mathrm{Pa}$ and $2.4\times10^2\,\mathrm{Pa}$.
The energy loss between the alpha source and the CCD
 depends on the amount of material between them. 
Here, the energy of the alpha particle at the CCD varied
 between $0.001\,\mathrm{MeV}$ and $\,\mathrm{4.75 MeV}$.
The energy of the event, shown in Fig.\,\ref{fig:cluster}, was
 calculated to be $2.0\,\mathrm{MeV}$.

\begin{figure}[htbp]
\begin{center}
\includegraphics[width=0.9\columnwidth]{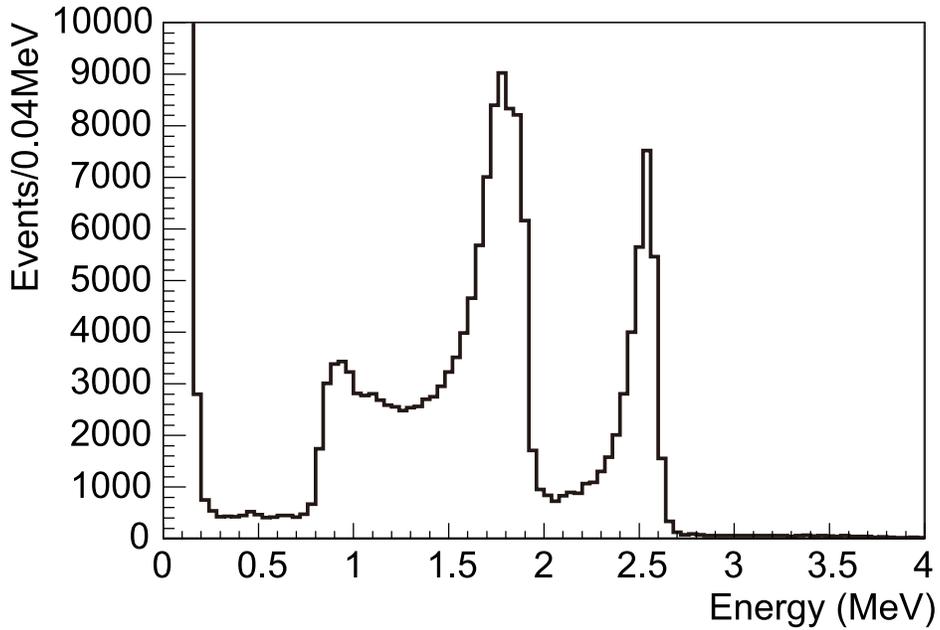}
\caption{Energy spectrum observed with the CCD coated with $^6$Li.}
\label{fig:energy}
\end{center}
\end{figure}

Fig.\,\ref{fig:energy} shows the energy spectrum
 obtained at the MINE2 cold neutron beam line.
It clearly shows the two peaks around $1.8\,\mathrm{MeV}$ and $2.5\,\mathrm{MeV}$, 
 which correspond to the alpha particles ($2.05\,\mathrm{MeV}$)
 and the tritons ($2.73\,\mathrm{MeV}$), respectively.
The observed energy is slightly smaller than the expected value
 because the charged particles lose a fraction of their energy
 in the converter and the insensitive volume of the CCD.
Fig.\,\ref{fig:energy} shows another peak around $0.9\,\mathrm{MeV}$,
 which decreases gradually with increasing energy,
 but has a sharp edge on the lower energy side.
This spectral shape is caused by tritons that are energetic
 enough such that their range is longer than the thickness of the CCD sensitive volume.
Supposing that the tritons stopping power (energy loss rate) is constant
 throughout the CCD,
 the energy deposited inside of the CCD sensitive volume, $E$,
 is a function of the incident angle $\theta$ as $E=t\rho\epsilon/\cos\theta$,
 where $t$ is the thickness of the CCD sensitive volume,
 $\rho$ is the density of the silicon, and
 $\epsilon$ is the stopping power ($\epsilon\equiv-dE/dx$).
Since the secondary particles are emitted isotropically,
 the number of events $dN(\theta)$ in between $\theta$ and $\theta+d\theta$ is
 proportional to $2\pi\sin\theta d\theta$.
The shape of the energy spectrum will be
 like $dN/dE \propto E^{-2}$; an inverse-square shape that should have upper and lower limits.
For incident particles with an initial energy $E_0$,
 all events with an incident angle larger than $\cos^{-1}(t\rho\epsilon/E_0)$
 will be accumulated at $E_0$, resulting in a peak.
The lower edge of the deposited energy corresponds to the vertically incident particles.
The deposited energy is equal to $t\rho\epsilon$,
 which amounts to about $0.9\,\mathrm{MeV}$ in our experiment.

The distribution of the energy lost
 inside the converter and the CCD insensitive volume
 will have the same inverse-square shape
 as the distribution inside the CCD sensitive volume described above.
Therefore,
 each peak, corresponding to the alpha particles and tritons that stop inside the CCD,
 has a tail proportional to $(E_0-E)^{-2}$ on the low energy side. 
The shape of this tail is blurred since the distance between 
 the neutron conversion point and the CCD surface can vary
 between zero and the full thickness of the converter.

\subsection{Position measurement}
\label{subsec:position}
The electrons spread out over the pixels, as shown in Fig.\,\ref{fig:cluster},
 due to diffusion inside the silicon.
The barycenter of the charge gives a finer hit-point
 than the pixel size.

\begin{figure}[htbp]
\begin{center}
\includegraphics[width=0.9\columnwidth]{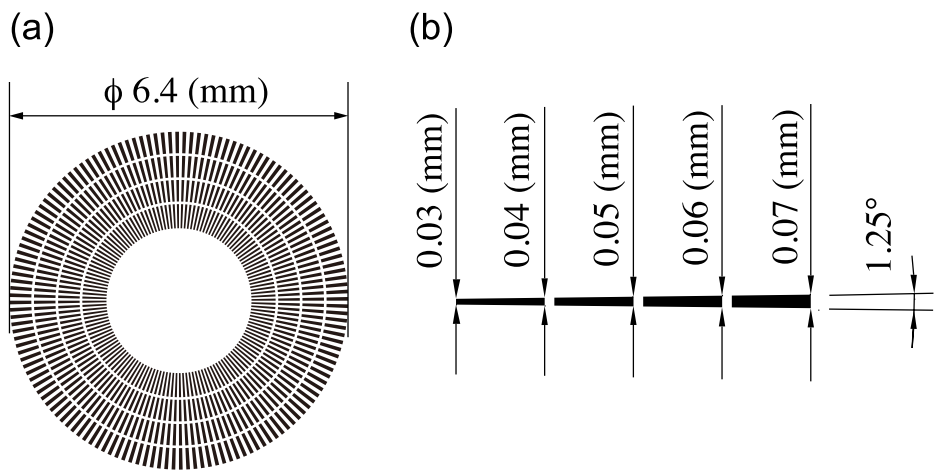}
\caption{(a) The reference pattern used for the resolution measurement, and (b)
 a close-up view of one ``spoke.''
}
\label{fig:spoke}
\end{center}
\end{figure}

A reference pattern was produced with a gadolinium foil to estimate the hit-position resolution for neutrons.
The thickness of the foil is 25~$\mu$m,
 which is thick enough to fully absorb cold neutrons.
The 144 fine spoke-like slits shown in Fig.\,\ref{fig:spoke}(a) were produced
 with an excimer laser by Laserx CO.LTD.
Each of the spoke-like slits has four trapezoidal holes,
 as shown in Fig.\,\ref{fig:spoke}(b).
The reference pattern was placed just in front of the $\mathrm{^6Li}$-coated CCD.
The distance between the pattern and the CCD was as small as $150\,\mu\mathrm{m}$.
The CCD and reference pattern
 were exposed to cold neutrons 
 at the MINE2 beam line of the research reactor JRR-3M.
The pattern was then projected onto the CCD.

\begin{figure}[htbp]
\begin{center}
\includegraphics[width=0.9\columnwidth]{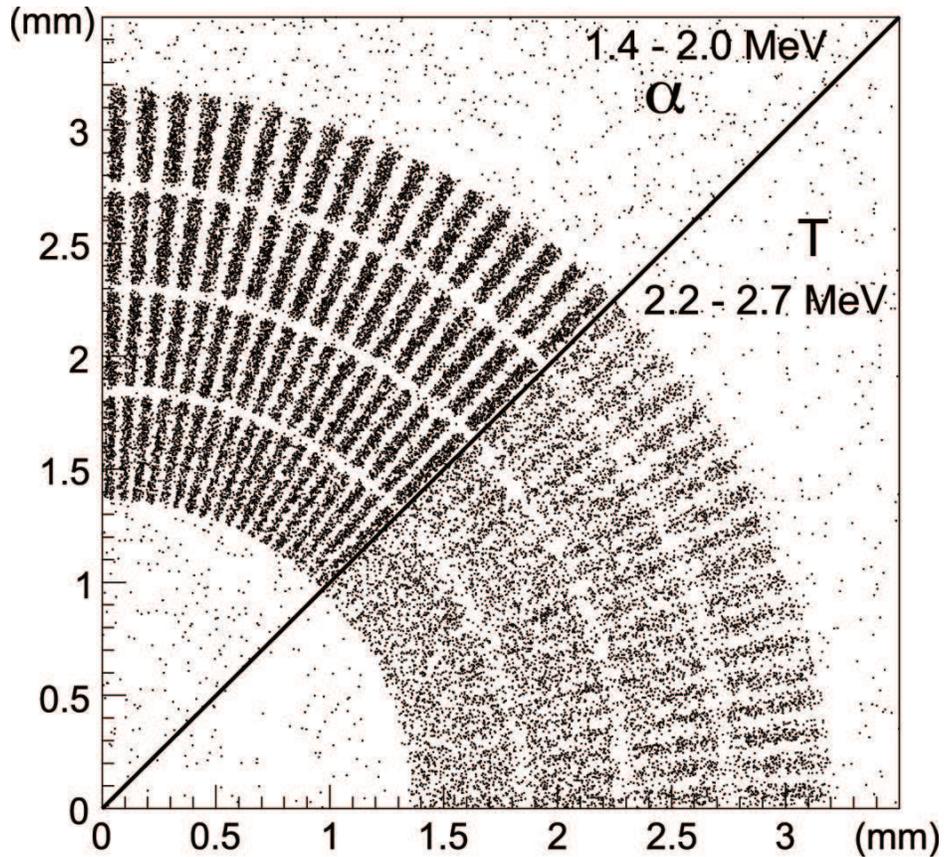}
\caption{The hit-position distribution of the alpha particles and the tritons.}
\label{fig:scatter}
\end{center}
\end{figure}

Fig.\,\ref{fig:scatter} shows hit-position distributions
 for the alpha particles and tritons. These distributions are
 estimated by calculating the barycenter of the deposited charge. 
In order to select alpha particles and tritons,
 the amount of energy deposited in each event was required
 to be in the range of $1.4\,\mathrm{MeV}$ -- $2.0\,\mathrm{MeV}$
 and $2.2\,\mathrm{MeV}$ -- $2.7\,\mathrm{MeV}$, respectively.
All of the events are plotted in an octant area,
 assuming that the event distribution has an 8-fold rotational symmetry.
For the tritons, 
 we were not able to see a pattern along the innermost part of the fan shape. 
This is because the tritons have a long range inside of the CCD.
On the other hand,
 a clear pattern can be seen for the alpha particles.
The hit-position distribution for alpha particles suggests that
 a spatial resolution of better than $30\,\mu\mathrm{m}$ was achieved. 

\begin{figure}[htbp]
\begin{center}
\includegraphics[width=0.9\columnwidth]{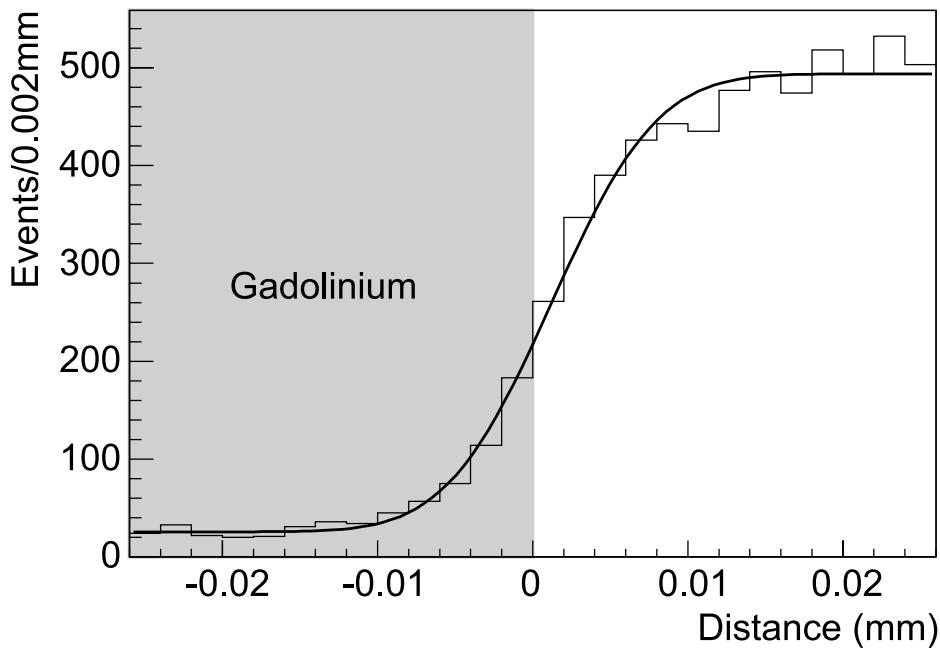}
\caption{The distance between the hit-position and the edge of the slit.}
\label{fig:reso}
\end{center}
\end{figure}

In order to estimate the spatial resolution more precisely,
 the distances between the calculated hit-positions and the edge of the slit was examined.
Fig.\,\ref{fig:reso} shows the distribution of these distances for alpha particles.
This distribution can be expressed as, 
\begin{equation}
f(x)=p_1 \cdot \mathrm{erf} \left( \frac{x}{\sqrt{2}\sigma}+p_2\right)+p_3,
\end{equation}
where $x$ is the distance, $\sigma$ corresponds to the spatial resolution,
and $p_1$, $p_2$ and $p_3$ are fitting parameters.
The estimated spatial resolution was
 $5.3 \pm 0.3\,\mu\mathrm{m}$ for alpha particles.
This is 4.5 times smaller than the CCD pixel size.

Our examination of the reference pattern via optical microscopy shows that
 the slits in the reference pattern were not perfectly straight. This is
 probably because the gadolinium foil was too thick
 to make slits as narrow as $30\,\mu\mathrm{m}$.
The variations from a straight slit edge were observed to be 1--$2\,\mu\mathrm{m}$.
The slit pattern had other imperfections as well:
 the slit walls were not strictly perpendicular to the face surface.
Our estimation of the detector resolution would be slightly worsened by both of these effects. 
 
\section{Application to ultracold neutron experiment}
A neutron's energy is quantized
 when it is bound in the earth's gravitational field. 
Given this effect, we expect to observe a modulation in the vertical distribution of neutrons in terrestrial experiments.
The scale of this modulation is about $6\,\mu\mathrm{m}$, so a magnification system is needed to observe the modulation with a CCD-based neutron detector whose spatial resolution is around $5\,\mu\mathrm{m}$.

\begin{figure}[htbp]
\begin{center}
\includegraphics[width=0.9\columnwidth]{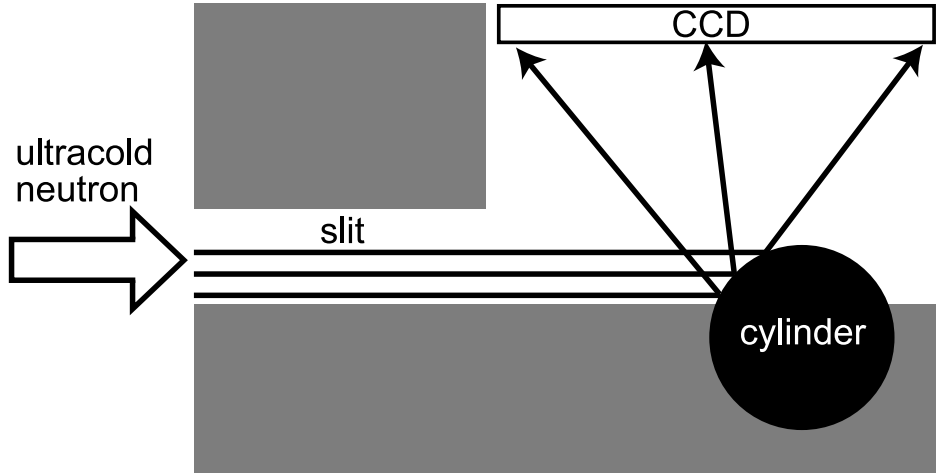}
\caption{An experimental setup (not to scale).}
\label{fig:setup}
\end{center}
\end{figure}

A cylinder is used as a convex mirror
 to magnify the small modulation pattern of the neutron distribution,
 as shown in Fig.\,\ref{fig:setup}.
The cylinder converts horizontal velocity into vertical velocity,
 after which the neutrons escape from the earth's gravitational field
 and follow their classical trajectories.
 
\begin{figure}[htbp]
\begin{center}
\includegraphics[width=0.9\columnwidth]{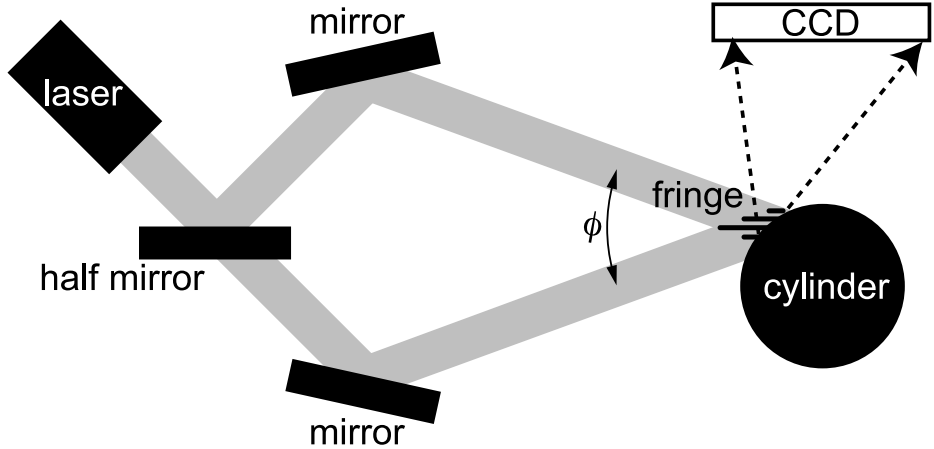}
\caption{The layout of the optical components 
 used in the magnification demonstration. (not to scale)}
\label{fig:optics}
\end{center}
\end{figure}

The magnification was demonstrated
 by observing a magnified laser interference fringe pattern with the CCD.
When two coherent laser beams intersect at an angle $\phi$,
 they produce a pattern of interference fringes.
The fringe spacing is given by $\lambda/2\sin(\phi/2)$,
 where $\lambda$ is the wavelength of the laser light \cite{Shintake:1991jb,Shintake:1992yu}.
This fringe pattern was magnified with the cylinder and then observed with the CCD.
A layout of the optical components we used is shown in Fig.\,\ref{fig:optics}.
In our demonstration,
 a green Nd:YAG laser ($\lambda$=532~nm) was used
 and the angle of intersection was measured to be $0.0722\pm0.0003\,\mathrm{radian}$,
 resulting in a fringe spacing of $7.37 \pm0.03\,\mu\mathrm{m}$.
The diameter of the cylinder
 was $1.5\,\mathrm{mm}$, 
 and the distance between the center of the cylinder and the CCD was
 designed to be $11.25\,\mathrm{mm}$.

\begin{figure}[htbp]
\begin{center}
\includegraphics[width=0.9\columnwidth]{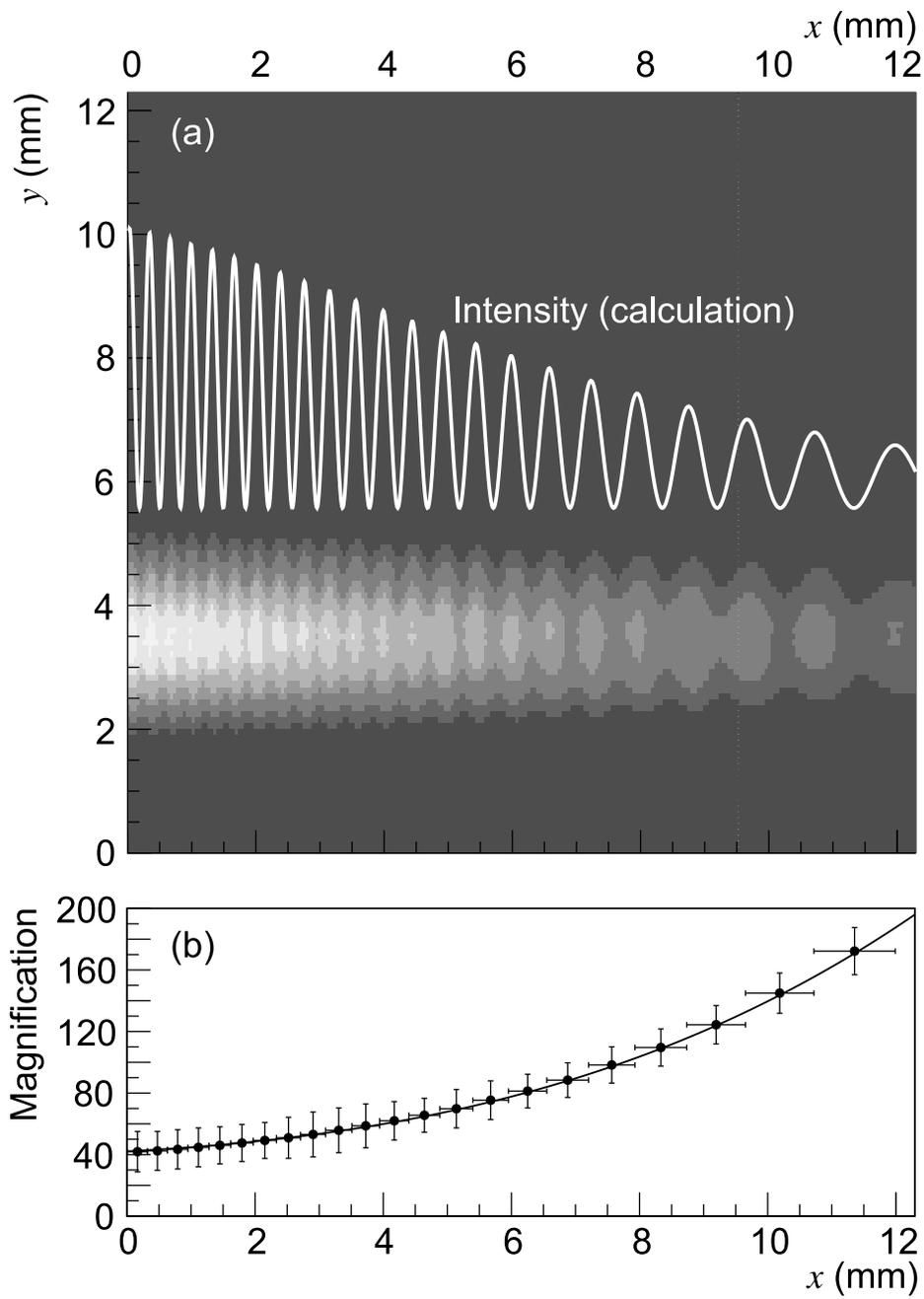}
\caption{(a) Magnified fringe pattern observed with the CCD, and
(b) the obtained magnification power.
The calculated intensity is superimposed as a white line in (a).}
\label{fig:magdemo}
\end{center}
\end{figure}

Fig.\,\ref{fig:magdemo}(a) shows the fringe pattern observed with the CCD.
The observed intensity distribution was reproduced in a calculation,
 shown in the figure by the white curve.
The magnifying power was calculated
 to be 40 -- 200
 from the observed fringe spacing on the CCD,
 as shown in Fig.\,{\ref{fig:magdemo}}(b).
The error
 in the estimated magnifying power
 is mainly due to uncertainty
 in determining the peak positions of the fringe pattern.
The horizontal error bars show the distance between the adjacent fringes.
Fig.\,\ref{fig:magdemo} shows that the magnification system functions as designed.
These results demonstrate that
 with this simple magnification system, 
 where a single cylinder is used as a convex mirror,
  an effective spatial resolution of $0.1\,\mu\mathrm{m}$
 can be achieved
 combined with the CCD-based neutron detector.
 
In order to precisely measure the position of the ultracold neutrons, we need to further optimize the material of the cylinder, as well as
 its diameter and distance to CCD.
In order to fully understand the magnification system
 and precisely estimate the neutron distribution shown by the CCD,
 it would be needed to calculate the evolution of the neutron wave function
 for the specific experimental situation of our measurement.

\section{Summary}
We developed a CCD-based pixel sensor for measuring neutrons.
A sufficient energy measurement capability and
 a spatial resolution of $5.3~\mu$m were demonstrated
 using the cold neutrons at the MINE2 beam line of the JRR-3M research reactor.
We propose that this sensor can be used to observe the quantum states of neutrons
 bound in the earth's gravitational field.
The neutron density modulation pattern can be stretched by
 two orders of magnitude
 using a simple magnification system
 composed of a cylindrical neutron reflector of $1.5\,\mathrm{mm}$ in diameter.
We have demonstrated
 that the fine structure of the neutron density modulation can be clearly observed
 using the proposed CCD-based detector
 coupled to the cylinder-based magnification system.

\section{Acknowledgements}
We are indebted to Prof. M.~Hino and Dr. M.~Kitaguchi 
 from Kyoto University Research Reactor Institute (KURRI), and 
Prof. H.M.~Shimizu and Prof. Y.~Higashi from High Energy Accelerator Research Organization (KEK)
for their technical support and encouragement. 
We thank Dr. Y.~Kamiya, Mr. H.~Okawa and Mr. N.~Mori
 from The University of Tokyo for their support.
This work was partially supported by
 KAKENHI (18340056) and
 Grant-in-Aid for JSPS Fellows (19$\cdot$4404).


\begin{thebibliography}{00}

\bibitem{Nesvizhevsky:2002ef}
V.~V. Nesvizhevsky, et~al.,
Nature 415 (2002) 297--299.

\bibitem{Nesvizhevsky:2003ww}
V.~V. Nesvizhevsky, et~al.,
Phys. Rev. D67 (2003) 102002.

\bibitem{Nesvizhevsky:2005ss}
V.~V. Nesvizhevsky, et~al.,
Eur. Phys. J. C40 (2005) 479--491.

\bibitem{Ebisawa:1994ru}
T.~Ebisawa, et~al.,
Nucl. Inst. Methods A350 (1994) 296--299.

\bibitem{KUR}
S.~Tasaki, et~al.,
Nucl. Inst. Methods A355 (1995) 501--505.

\bibitem{Shintake:1991jb}
T.~Shintake, in: Proc. of the 8th Symp. on
  Accel. Sci. and Technol., Ionics Pub., Tokyo, Japan, 1991, pp.
  290--292.

\bibitem{Shintake:1992yu}
T.~Shintake, et~al.,
Int. J. Mod. Phys. Proc. Suppl. 2A (1993) 215--218.

\end{thebibliography}
\end{document}